# Sensors for healthcare: Would you want them in your home?

**Alison Burrows**
SPHERE IRC
Merchant Venturers Building
Woodland Road
Bristol BS8 1UB, UK
alison.burrows@bristol.ac.uk

**Rachael Gooberman-Hill**
Bristol Implant Research
Avon Orthopaedic Centre
Southmead Hospital
Bristol BS10 5NB, UK
r.gooberman-hill@bristol.ac.uk

**Ian Craddock**
SPHERE IRC
Merchant Venturers Building
Woodland Road
Bristol BS8 1UB, UK
ian.craddock@bristol.ac.uk

**David Coyle**
Dept. of Computer Science
University of Bristol
Woodland Road
Bristol BS8 1UB, UK
david.coyle@bristol.ac.uk

## Abstract
This paper describes some of the challenges set within SPHERE, a large-scale Interdisciplinary Research Collaboration that aims to develop sensor systems to monitor people's health and wellbeing in the home. In particular we discuss the dual task facing the User-Centered Design research group, to ensure the development of inclusive and desirable domestic healthcare technology. On the one hand, we seek to gain a rich understanding of the many envisaged users of the SPHERE system. On the other hand, for the user experience requirements to be translated into tangible outputs, it is crucial that we effectively communicate these findings to the broader team of SPHERE engineers and computer scientists.

## ACM Classification Keywords
H.5.0. Information interfaces and presentation: General

## Introduction
SPHERE (Sensor Platform for HEalthcare in a Residential Environment) is an Interdisciplinary Research Collaboration (IRC), with the vision of establishing a common platform of non-medical/environmental sensors to impact a variety of healthcare needs. These sensors can be categorized as:

- Indirect, for example detecting human behavior through home energy use;
- Remote, specifically detecting human behavior through video monitoring;
- On-body, which includes using sensors situated on the person for monitoring purposes as well as energy harvesting and management.

As researchers, this is an exciting challenge. And it is not difficult to conceive of a user group or even an older relative who could benefit from this type of technology. However, the prospect of stepping into the user's shoes, of personally allowing this technology into homes and onto bodies is met with apprehension. On this point, Fulton Suri observes [1]:

"On the one hand, many design problems arise when we assume that everyone else is just like us. Poor design is often the result of [this] assumption […]. On the other hand, many problems arise when we think of other people as so different from ourselves that we think of them as 'them'."

It is therefore not surprising that, in spite of the strong social and economic case for its adoption, healthcare technology has yet to realize its potential at scale. In fact, the EU-funded EFFORT project produced a comprehensive report on technological care interventions for older citizens highlighting the existence of a fundamental socio-technical gap in existing systems [2].

The SPHERE IRC, which is in the early stages of its five year timescale, envisages a clinically effective system. Likewise, it is aims to produce outputs that are meaningful and desirable. In order to achieve this, it intends to establish early and sustained involvement with a range of stakeholders such as domestic users, care givers, clinicians and social workers.

**User-Centered Design approach**

Many studies of technology-assisted healthcare are conducted in living lab situations (for an overview, see [3]). While these labs allow detailed information to be gathered on systems' clinical effectiveness, they represent a compromise in terms of the contextual complexity of healthcare practices. This has had implications for the telehealth and telecare literature, which for the most part has failed to address the fact that self-management of illness is neither rational, nor simply a matter of processing information [4]. If healthcare technologies are to become embedded into people's everyday life, developers need to take a holistic and empathic view of their target users.

For SPHERE, this requires a broad contextual investigation of people's current healthcare practices, alongside their experiences with technology. Given the scope of this IRC, the contexts of use have been defined as the Self, the Home and the Community (see Figure 1). In a first phase, this will entail ethnographic studies with 15 to 20 households.

As the User-Centered Design (UCD) work package on SPHERE, we will draw on skills developed through prior user-sensitive research (e.g. [5], [6]). Our studies will have a *participatory mindset* since we see people as experts of their own experience and, therefore, as uniquely qualified to contribute to the design process [7]. In practice, this means using a combination of methods and tools that empower people to share rich experiential knowledge, including cultural probes [8] and technology probes [9].

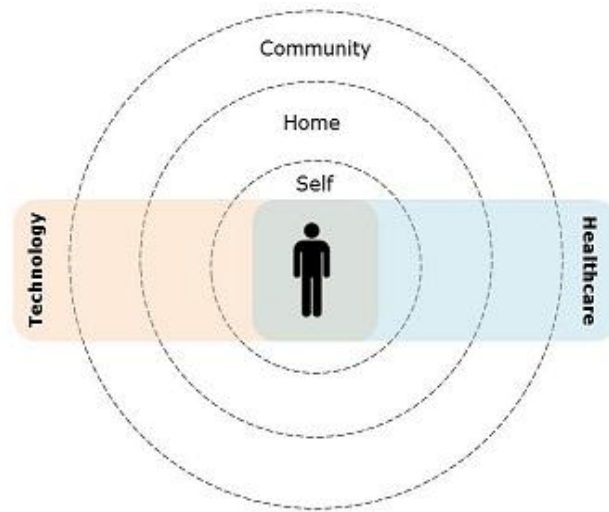

**Figure 1.** UCD research contexts for SPHERE.

## Keeping the empathy torch alive

As researchers who will have first-hand and prolonged contact with end users, the UCD group has an advantage in establishing empathic engagement. An important challenge we then face is to foster a similar understanding of the users in the engineers and computer scientists, who may be more technology-focused. Typically, multidisciplinary teams such as SPHERE need also to overcome barriers resulting from the use of discipline-specific jargon and time constraints to achieving a deep understanding of intended users. There are a number of rapid, immersive techniques that can be used to provoke reflective thinking from another person's standpoint. For instance, empathic modeling is a method that aims to simulate disabilities through the use of props, thus encouraging people to think about coping strategies and adaptation techniques [10].

However, the physical performance of products is just one aspect of the user experience, which comprises of an array of intangible components including motivation, emotions and aspirations. Effective means for communicating this type of rich experiential information contain visual material, subjective information, unfiltered information, and stories [11]. Personas are one method that can condense large amounts of information about how people behave, what they want to achieve and what they fear as users. The term *persona* was popularized by Alan Cooper [12], but has since become widespread as a user-centered method amongst designers and companies, notably Microsoft, Philips and the BBC. Personas provide a manageable medium for creating empathy between designers and the real people who will be using their products and services [13]. They therefore humanize crucial usability and user experience data, maintaining a UCD focus throughout the design process. From previous research experience, we found that conducting workshops with data-driven personas was effective and engaging [14]. Moreover, designers felt that using personas stimulated their design thinking and provided them with inspiration.

## Acknowledgements
The authors gratefully acknowledge funding for the SPHERE IRC from the UK Engineering and Physical Sciences Research Council (EPSRC).